\pdfoutput=1
\documentclass[preprint2, times]{aastex6}

\usepackage{natbib}
\usepackage{amsmath}
\usepackage{amssymb}
\usepackage{url}

\renewcommand{\vec}[1]{ {\mathbf #1} }

\newcommand{\grad}{ {\bf \nabla } }

\newcommand{\Eq}{{Equation}}

\newcommand{\Fig}{{Figure}}

\newcommand{\dive}{\nabla\cdot}

\providecommand{\dodoi}[1]{doi:~\href{http://doi.org/#1}{\nolinkurl{#1}}}
\providecommand{\url}[1]{\href{#1}{#1}}
\providecommand{\doeprint}[1]{\href{http://ascl.net/#1}{\nolinkurl{http://ascl.net/#1}}}
\providecommand{\doarXiv}[1]{\href{https://arxiv.org/abs/#1}{\nolinkurl{https://arxiv.org/abs/#1}}}
\newcommand{\gradxy}{\nabla_{\perp}}

\begin{document}

\title{MHD Modeling of Solar Coronal Magnetic
  Evolution Driven by Photospheric Flow}

\author{Chaowei Jiang\altaffilmark{1}, Xinkai Bian\altaffilmark{1},
  Tingting Sun\altaffilmark{1}, Xueshang Feng\altaffilmark{1}}

\altaffiltext{1}{Institute of Space Science and Applied Technology,
  Harbin Institute of Technology, Shenzhen 518055, China; chaowei@hit.edu.cn
}

\begin{abstract}
  It is well known that magnetic fields dominate the dynamics in the
  solar corona, and new generation of numerical modelling of the
  evolution of coronal magnetic fields, as featured with boundary
  conditions driven directly by observation data, are being developed.
  This paper describes a new approach of data-driven
  magnetohydrodynamic (MHD) simulation of solar active region (AR)
  magnetic field evolution, which is for the first time that a data-driven full-MHD model utilizes directly the photospheric velocity field from DAVE4VM. We constructed a {well-established} MHD equilibrium
  based on a single vector magnetogram by employing an MHD-relaxation
  approach with sufficiently small kinetic viscosity, and used this MHD equilibrium as
  the initial conditions for subsequent data-driven evolution. Then we
  derived the photospheric surface flows from a time series of
  observed magentograms based on the DAVE4VM method. The surface flows
  are finally inputted in time sequence to the bottom boundary of the
  MHD model to self-consistently update the magnetic field at every
  time step by solving directly the magnetic induction equation at the
  bottom boundary. We applied this data-driven model to study the
  magnetic field evolution of AR~12158 with SDO/HMI vector
  magnetograms. Our model reproduced a quasi-static stress of the
  field lines through mainly the rotational flow of the AR's leading
  sunspot, which makes the core field lines to form a coherent S shape
  consistent with the sigmoid structure as seen in the SDO/AIA
  images. The total magnetic energy obtained in the simulation matches
  closely the accumulated magnetic energy as calculated directly from
  the original vector magnetogram with the DAVE4VM derived flow
  field. Such a data-driven model will be used to study how the coronal
  field, as driven by the slow photospheric motions, reaches a
  unstable state and runs into eruptions.
\end{abstract}


\section{Introduction}
\label{sec:intro}

Magnetic fields dominate the dynamics in the Sun's upper atmosphere,
the solar corona. On the solar surface, i.e., the photosphere,
magnetic fields are seen to change continuously; magnetic flux
emergence brings new flux from the solar interior into the atmosphere,
and meanwhile the flux is advected and dispersed by surface motions
such as granulation, differential rotation and meridional
circulation. Consequently, the coronal field evolves in response to
(or driven by) the changing of the photospheric field, and thus
complex dynamics occur ubiquitously in the corona, including the
interaction of newly emerging field with the pre-existing one,
twisting and shearing of the magnetic arcade fields, magnetic
reconnection, and magnetic explosions which are manifested as flares
and coronal mass ejections (CMEs).

As we are still not able to measure directly the three-dimensional
(3D) coronal magnetic fields, numerical modelling has long been
employed to reconstruct or simulate the coronal magnetic fields based
on different assumption of the magnetohydrodynamic (MHD) equations,
such as the most-frequently used force-free field
model~\citep{Wiegelmann2012solar}, which has been developed for
over four decades. However, the force-free field
assumption is only valid for the equilibrium of the corona, and it
cannot be used to follow a continual and dynamic evolution of the
magnetic fields. In recent years, with vector magnetogram data in the photosphere
measured routinely with high resolution and cadence, data-driven
modelling is becoming a viable tool to study coronal magnetic field
evolution, which can self-consistently describe both the quasi-static
and the dynamic evolution phases~\citep[e.g.,][]{Wu2006, FengX2012,
  Cheung2012, Jiang2016NC, Leake2017, Inoue2018, Hayashi2019,
  GuoY2019, Pomoell2019}. Still, due to the limited constraint from
observation, data-driven models are developed with very different
settings from each other~\citep{Toriumi2020}. For simplicity, some used the
magneto-frictional model~\citep{Cheung2012, Pomoell2019}, in which the
Lorentz force is balanced by a fictional plasma friction force. As
such, the magnetic field evolves mainly in a quasi-static way~\citep{Yang1986},
although in some case an eruption can be reproduced but it evolves in
a much slower rate than the realistic one~\citep{Pomoell2019}, because the dynamic is
strongly reduced by the frictional force. Some used the so-called
zero-$\beta$ model~\citep{Inoue2016Review, Inoue2018, GuoY2019}, in which the gas
pressure and gravity are neglected. The zero-$\beta$ model might fail
when there is fast reconnection in the field, in which the thermal
pressure could play an important role in the dynamics in weak-field region of magnetic
field dissipation. Therefore, it is more realistic and {accurate} to
solve the full MHD equations to deal with the nonlinear interaction of magnetic fields with plasma.

To drive the full MHD model, one needs to specify all the eight
variables (namely plasma density, temperature, and three components of
velocity and magnetic field, respectively) in a self-consistent way at
the lower boundary. Previously, with only the magnetic field obtained
from observations, we employed the projected-characteristic
method~\citep{Nakagawa1980, Nakagawa1987, Wu1987, Hayashi2005, Wu2006}
to specify the other variables according to the information of
characteristics based on the wave-decomposition principle of the full
MHD system~\citep[e.g.,][]{Jiang2016ApJ,Jiang2016NC}. Specifically,
the full MHD equations are a hyperbolic system that can be
eigen-decomposed into a set of characteristic wave equations (i.e.,
compatibility equations), which are independent with each other; on
the boundary, these waves may propagate inward or outward of the
computational domain because the wave speeds (the eigenvalues) of both
signs generally exist at the boundary. If the wave goes out of the
computational domain through the boundary, it carries information
from the inner grid to the boundary, and thus the corresponding compatibility
equation should be used to constrain the variables on the boundary
surface. Thus if there are five waves going out, and with three
components of magnetic field specified by observed data, the eight
variables are fully determined. Ideally and in principle, the
projected-characteristic method is the most self-consistent one with
inputting data that is partially available (for example, in our case,
only the data of magnetic field).  However, such conditions, i.e.,
exactly five waves going out, are often not satisfied in the whole
boundary and other assumptions are still necessary. Even worse, in
areas near the magnetic polarity inversion line (PIL) where the normal magnetic field component is small,
the Alfv{\'e}n wave information goes mainly in the transverse
direction rather than the normal one, and the projected-characteristic
approach may fail. Another shortcoming of the method lies in its
difficulty in code implementation, which needs to perform
eigen-decomposition and solve a linear system of the compatibility
equations on every grid point on the boundary to recover the primitive
variables from the characteristic ones.


In this paper we test another way of specifying the bottom boundary
conditions, which uses the surface velocity derived by the DAVE4VM
method~\citep{Schuck2008} to drive the MHD model. The differential
affine velocity estimator (DAVE) was first developed for estimating
velocities from line-of-sight magnetograms, and was then modified to
directly incorporate horizontal magnetic fields to produce a
differential affine velocity estimator for vector magnetograms
(thus called DAVE4VM). It is generally accepted that the coronal magnetic fields are
line tied in the dense photosphere and are advected passively by the surface
motions of the photosphere, such as the shearing, rotational, and converging flows~\citep{Priest2002}.
Thus, with the surface velocity in hand, we can solve the
magnetic induction equation on the bottom boundary to update
self-consistently the magnetic field to implement such a line-tying boundary condition,
which is a much simpler way than
using the projected-characteristic method. To illustrate the approach,
we take the solar AR 12158 as an example to simulate its two-day
evolution from 2014 September 8 to 10 during its passage of the solar
disk. This AR is of interest since it produced an X1.6 eruptive flare
accompanied with a fast CME with speed of $\sim 1300$~km~s$^{-1}$,
which is well documented in the literature~\citep{Vemareddy2016,
  ZhaoJ2016, ZhouG2016}. The AR is well isolated from neighboring ARs,
thus is suited for our modelling focused on a single AR. Within a few days prior to this
major eruption, the AR developed from a weakly sheared magnetic arcade
into a distinct sigmoidal configuration, indicating a
continual injection of non-potential magnetic energy into the corona through the
photospheric surface motion. Indeed, prior to the eruption, the major
sunspot of the AR showed a significant rotation of over $200^{\circ}$
in five days~\citep{Vemareddy2016}. Based on the sigmoidal hot coronal loop
seen immediately before the flare, many authors have interpreted it as a
pre-existing magnetic flux rope which erupted and resulted in the flare
and CME~\citep[e.g.,][]{Vemareddy2016, ZhaoJ2016, ZhouG2016}, while
some nonlinear force-free field extrapolations appear not to support
this~\citep{DuanA2017, Lee2018}. Thus a fully data-driven MHD
simulation can provide valuable insight in addressing this issue by
following the dynamic evolution of the coronal magnetic field,
although the main objective of this paper is to describe the methods.

In the following we first describe our model equation in
Section~\ref{sec:model}. The data-driven simulation consists of three
steps, and first we constructed an MHD equilibrium based on a single
vector magnetogram observed for the start time of our simulation,
which is described in Section~\ref{initial}. Then in
Section~\ref{flow} we calculated the surface flow field using the
DAVE4VM code with the time series of vector magnetograms. Finally, we
input the flow field in the model to drive the evolution of the MHD
system, as described in Section~\ref{evol}. Summary and discussion are
given in Section~\ref{concl}.

\begin{table}[htbp]
  \centering
  \begin{tabular}{lll}
    \hline \hline
    Variable &  Expression & Value \\
    \hline
    Density        & $\rho_{s}$ =  $n m $ & $2.29 \times 10^{-15}$~g~cm$^{-3}$ \\
    Temperature    & $T_{s}$       & $1\times 10^{6}$~K \\
    Length         & $L_{s}$  = $16$~arcsec  &  $11.52$~Mm\\
    Pressure       & $p_{s}=2nk_{B}T_{s}$   &  $2.76\times 10^{-2}$~Pa \\
    Magnetic field & $B_{s}=\sqrt{\mu_{0}p_{s}}$ & $1.86$~G\\
    Velocity       & $v_{s}=\sqrt{p_s / \rho_{s} }$ & $110$~km~s$^{-1}$ \\
    Time           & $t_{s}=L_{s}/v_{s}$ & $105$~s \\
    Gravity        & $g_{s}=v_{s}/t_{s}$ & $1.05$~km~s$^{-2}$\\
    \hline
  \end{tabular}
  \caption{Parameters used for non-dimensionalization. $n$ is a
    typical value of electron number density in the corona given by
    $n=1\times 10^{9}$~cm$^{-3}$ and $m$ is the mean atomic mass.}
  \label{tab:table1}
\end{table}

\section{MHD Equations}
\label{sec:model}

We numerically solve the full MHD equations in a 3D Cartesian geometry
by an advanced conservation element and solution element (CESE)
method~\citep{Jiang2010}.  Before describing the model equations in
the code, it is necessary to specify the quantities used for
non-dimensionalization. Here we use typical values at the base of the
corona for non-dimensionalization as given in
Table~\ref{tab:table1}. In the rest of the paper all the variables and
quantities are written in non-dimensionalized form if not mentioned
specially.

In non-dimensionalized form, the full set of MHD equations are given as
\begin{eqnarray}
  \label{eq:MHD}
  \frac{\partial \rho}{\partial t}+\dive (\rho\vec v) =
  -\nu_{\rho}(\rho-\rho_0),\nonumber \\
  \rho\frac{D\mathbf{v}}{D t} = -\grad p+\vec J\times \vec B+\rho\vec
  g + \nabla\cdot(\nu\rho\nabla\mathbf{v}),\nonumber\\
  \frac{\partial \vec B}{\partial t} =
  \grad \times (\vec v \times \vec B), \nonumber\\
  \frac{\partial T}{\partial t}+\nabla\cdot (T\vec v) =
  (2-\gamma)T\nabla\cdot\vec v.
\end{eqnarray}
where $\vec J = \nabla \times \vec B$, $\nu$ is the kinetic viscosity,
and $\gamma$ is the adiabatic index.

Note that we artificially add a source term $-\nu_{\rho}(\rho-\rho_0)$
to the continuity equation, where $\rho_0$ is the density at the
initial time $t=0$ (or some prescribed form), and $\nu_{\rho}$ is a
prescribed coefficient. This term is used to avoid a ever-decreasing
of the density in the strong magnetic field region, which we often encounter
in the very low-$\beta$ simulation. It can maintain
the maximum Alfv{\'e}n speed in a reasonable level, which may
otherwise increase and make the iteration time step smaller and
smaller and the long-term simulation unmanageable. {Specifically, this source
term is a Newton relaxation of the density to its initial value by a
time scale of
\begin{equation}
 \tau_{\rho} = \frac{1}{\nu_{\rho}} = 20 \tau_{\rm A},
\end{equation}
where $\tau_{\rm A} = 1/v_{\rm A}$ is the Alfv{\'e}n time with length of 1 (the length unit) and the Alfv{\'e}n speed $v_{\rm A} = B/\sqrt{\rho}$. Thus it is
sufficiently large to avoid its influence on the fast dynamics of
Alfv{\'e}nic time scales. As a result, we used $ \nu_{\rho} = 0.05 v_{\rm A} $ in all the
simulation in this paper.}

No explicit resistivity is used in the magnetic induction equation,
but magnetic reconnection is still allowed through numerical diffusion
when a current layer is sufficiently narrow such that its width is
close to the grid resolution. In the energy (or temperature) equation,
we set $\gamma=1$ for simplicity, such that the energy equation
describes an isothermal process. \footnote{{Although in this case we can simply discard the energy equation by setting the temperature as a constant, we still keep the full set of equations in our code which can thus describe either the isothermal or adiabatic process by choose different value of $\gamma$.}} The kinetic viscosity $\nu$ will be given
with different values when needed, which is described in the
following sections.

\begin{figure*}[htbp]
  \centering
 \includegraphics[width=0.8\textwidth]{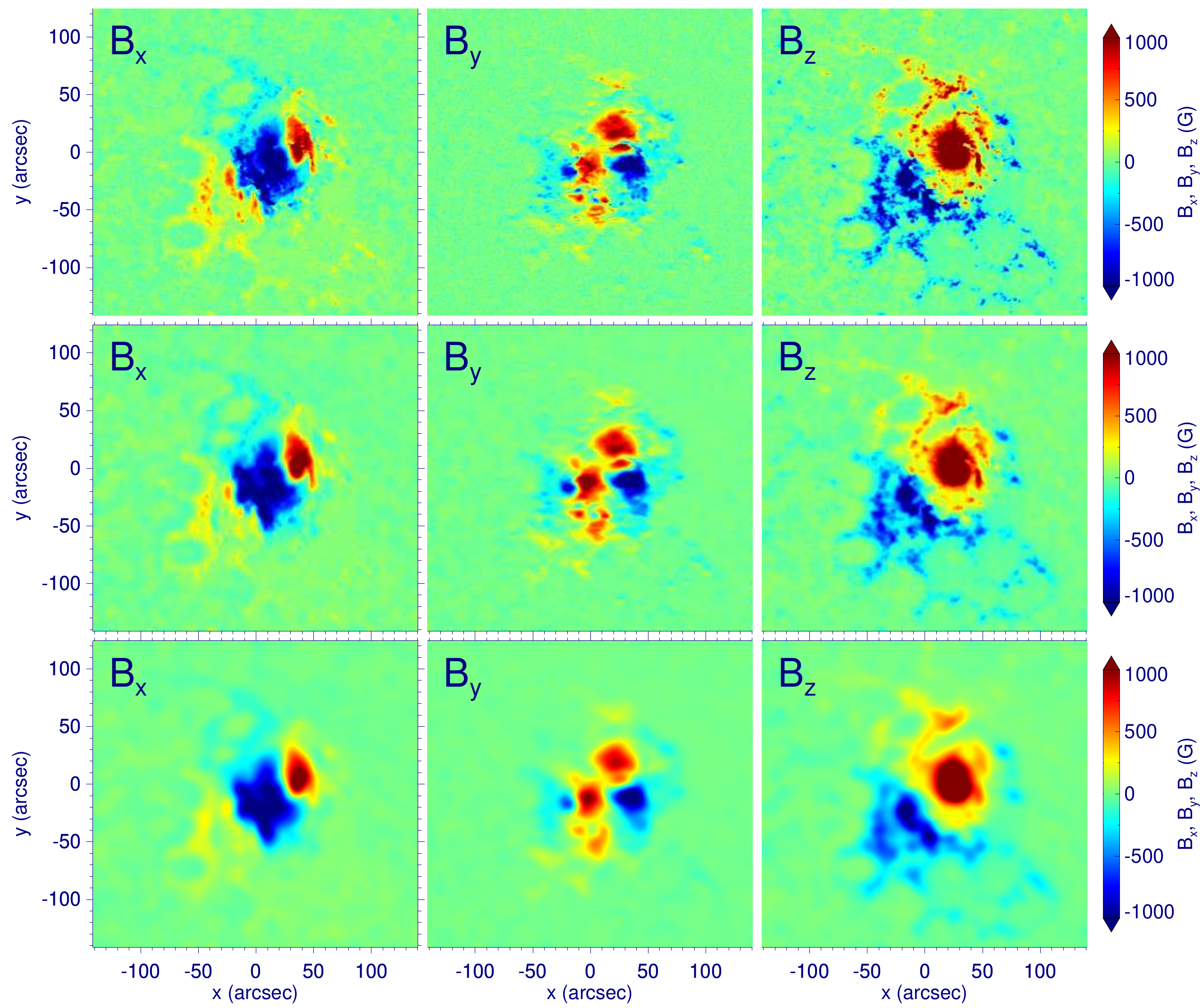}
 \caption{Comparison of the original HMI vector magnetogram
   for time of 00:00~UT on 2014 September 8 (top
   panels), the preprocessed one (middle) and the final smoothed
   one (bottom). From left to right are shown for magnetic field
   components $B_x$, $B_y$, and $B_z$, respectively.}
  \label{1}
\end{figure*}

\begin{figure}[htbp]
  \centering
 \includegraphics[width=0.45\textwidth]{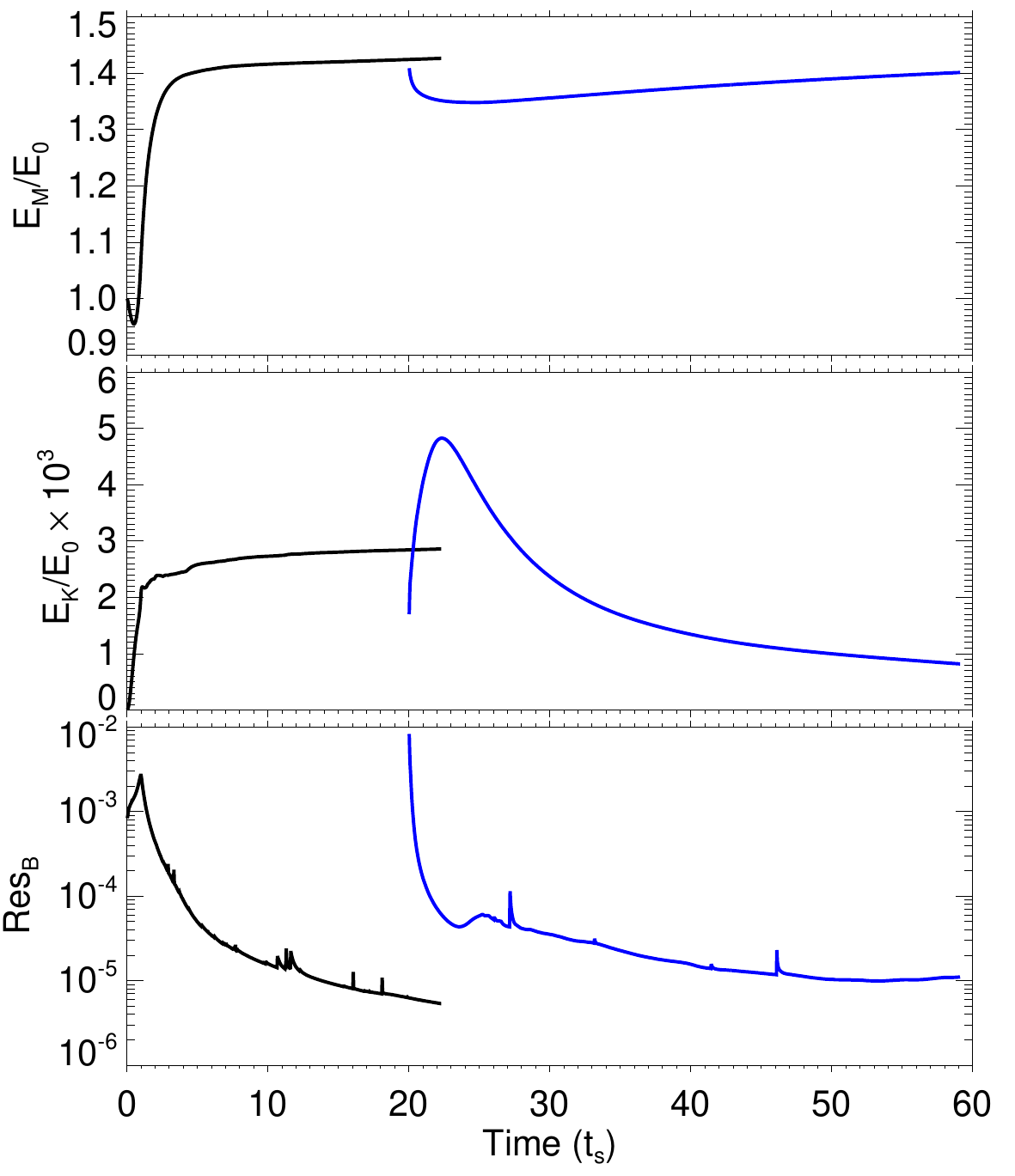}
 \caption{Evolutions of the total magnetic energy, kinetic energy, and
   residual of magnetic field in the process of the constructing the
   initial MHD equilibrium. The black curves show the results for the
   first evolution phase, and the blue ones show that for the second,
   `deeper' evolution phase.}
  \label{2}
\end{figure}

\begin{figure*}[htbp]
  \centering
 \includegraphics[width=0.7\textwidth]{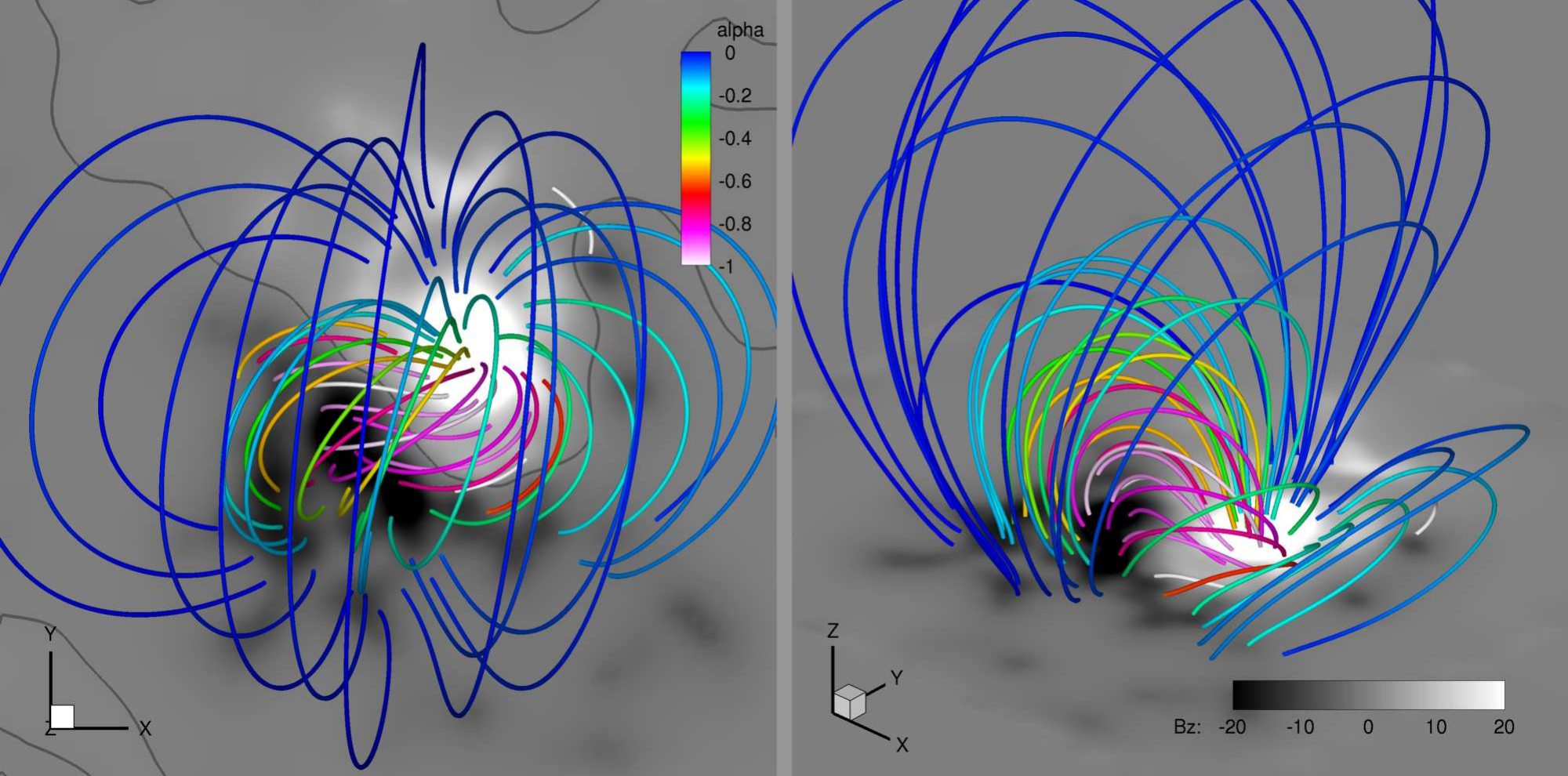}
 \caption{3D magnetic field of the MHD equilibrium based on the
   magnetogram taken for 00:00~UT on 2014 September 8. The thick lines
   are magnetic field lines and are pseudo-colored by value of the
   force-free parameter $\alpha$. The background is shown with the
   magnetic flux distribution (i.e., $B_z$) on the bottom surface of
   the model.}
  \label{3}
\end{figure*}

\section{Construction of an initial MHD equilibrium}
\label{initial}
We first constructed an initial MHD equilibrium based on a single
vector magnetogram taken for time of 00:00~UT on 2014 September 8 by
SDO/HMI. Such an equilibrium is assumed to exist when the corona is
not in the eruptive stage, and is crucial for starting our subsequent
data-driven evolution. The vector magnetogram is preprocessed to
{reduce} the Lorentz force and is further smoothed to filter out the
small-scale structures that are not sufficiently resolved in our
simulation. Reduction of the Lorentz force is helpful for reaching a
more force-free equilibrium state~\citep[e.g.,][]{Wiegelmann2004}, and
smoothing is also necessary to mimic the magnetic field at the coronal
base rather directly the photosphere, because the lower boundary of
our model is placed at the base of the corona~\citep{JiangC2020}. The
preprocessing is done using a code developed by~\citet{Jiang2014Prep},
which is originally designed for NLFFF extrapolation using the HMI
data~\citep{Jiang2013NLFFF}. {Specifically, the total Lorentz force and torque are quantified by two surface integrals associated with three components of the photospheric magnetic field, and an optimization method is employed to minimize these two functions by modifying the magnetic field components within margins of measurement errors.}
Then Gaussian smoothing with FWHM of
8~arcsec is applied to all the three components of the magnetic
field. \Fig~\ref{1} compares the original HMI vector magnetogram with
the preprocessed one as well as the final smoothed version. Using the
vertical component $B_z$ of this preprocessed and smoothed
magnetogram, a potential field is extrapolated and used as the initial
condition of the magnetic field in the MHD model.

In addition to the magnetic field, an initial plasma as the
background atmosphere is also needed to start the MHD simulation. We
used an isothermal plasma in hydrostatic equilibrium. It is stratified
by solar gravity with a density $\rho=1$ at the bottom and a uniform
temperature of $T=1$. Using this typical coronal plasma, we did not
directly input the magnetic field into the model but multiplies it
with a factor of $0.05$ such that the maximum of magnetic field
strength in normalized value is approximately $50\sim 100$ in the
model. If using the original values of magnetic field, its strength
(and the characteristic Alfv{\'e}n speed) near the lower surface is
too larger, which will be a too heavy burden on computation as the
time step of our simulation is limited by the CFL condition. With the
reduced magnetic field, we further modified the value of solar gravity
to avoid an unrealistic large plasma $\beta$ (and small Alfv{\'e}n speed) in the corona. This is
because if using the real number of the solar gravity
($g_{\odot} = 274$~m~s$^{-2}/g_{s} = 0.26$), it results in a pressure
scale height of $H_{p} = 3.8$, by which the plasma pressure and
density decay with height much slower than the magnetic field. With
the weak magnetic field strength we used, the plasma $\beta$ will
increase with height very fast to above $1$, which is not realistic in
the corona. To make the pressure (and density) decrease faster in the
lower corona, we modified the gravity as
\begin{equation}
  \label{eq:gravity}
  g = \frac{1.5H_{p}g_{\odot}}{(1 + 0.15z)^{2}}.
\end{equation}
By this, we got a plasma with $\beta<1$ mainly within $z<10$ and the
smallest value is $5\times 10^{-3}$.

Then we input the transverse field of the smoothed magnetogram to the
model. This is done by modifying the transverse field on the bottom
boundary incrementally using linear extrapolation from the potential
field to the vector magetogram in time with a duration of 1 $t_s$
until it matches the vector magnetogram. This will drive the coronal
magnetic field to evolve away from the initial potential state, since
the change of the transverse field will inject electric currents and
thus Lorentz forces, which drive motions in the computational
volume. Note that in this phase all other variables on the bottom
boundary are simply fixed, thus the velocity remaining zero. This is
somewhat un-physical since the Lorentz force will introduce nonzero
flows on the bottom boundary, but it provides a simple and `safe' way
(avoiding numerical instability) to bring the transverse magnetic
field into the model. Once the magnetic field on the bottom surface is
identical to that of the vector magnetogram, the system is then left
to relax to equilibrium with all the variables (including the magnetic
field) on the bottom boundary fixed. To avoid a too large velocity in
this phase such that the system can relax faster, we set the kinetic
viscosity coefficient as $\nu = 0.5\Delta x^{2}/\Delta t$, where
$\Delta x$ is the local grid spacing and $\Delta t$ the local time
step, determined by the CFL condition with the fastest magnetosonic
speed. Actually this is the largest viscosity one can use with given
grid size $\Delta x$ and time step $\Delta t$, because the CFL
condition for a purely diffusive equation with diffusion coefficient
$\nu$ requires $\Delta t \le 0.5 \Delta x^{2}/\nu$.

For the purpose of minimizing the numerical boundary influences
introduced by the side and top boundaries of the computational volume,
we used a sufficiently large box of
$(-32, -32, 0) < (x, y, z) < (32, 32, 64)$ embedding the field of
view of the magnetogram of $(-8.75, -8.25) < (x, y) < (8.75, 8.25)$.
The full computational volume is resolve by a non-uniform
block-structured grid with adaptive mesh refinement (AMR), in which
the highest and lowest resolution are
$\Delta x = \Delta y = \Delta z = 1/16$ (corresponding to $1$~arcsec
or 720~km, matching the resolution of the vector magnetogram) and
$1/2$, respectively. The AMR is controlled to resolve with the
smallest grids the regions of strong magnetic gradients and current
density. {The magnetic field outside of the area of the magnetograms on the lower boundary is given as zero for the vertical component and simply fixed as the potential field for the transverse components.} On the side and top boundaries, we fixed the plasma density,
temperature and velocity. The horizontal components of magnetic field
are linearly extrapolated from the inner points, while the normal
component is modified according to the divergence-free condition to
avoid any numerical magnetic divergence accumulated on the boundaries.

In \Fig~\ref{2}, the curves colored in black show the evolution of the
magnetic and kinetic energies integrated for the computational volume,
and the residual of the magnetic field of two consecutive time
steps which is defined as
\begin{equation}
  {\rm Res}_B = \sqrt{ \frac{1}{3} \sum_{\delta=x,y,z} \frac{\sum\nolimits_{i}
  \left( B_{i\delta}^{k}-B_{i\delta}^{k-1}\right) ^{2} }{\sum\nolimits_{i}\left( B_{i\delta}^{k} \right) ^{2} } },
\end{equation}
where the indices $k$ and $k-1$ refer to the two consecutive time
steps and $i$ goes through all the mesh points. It can be seen that
the magnetic energy increases sharply in a few time
units\footnote{Note that at the very beginning the magnetic energy
  actually decreases shortly for about $0.5~t_s$, which is unphysical
  as the potential field energy is in principle the lowest energy with
  a given magnetic flux distribution on the bottom boundary. Such an
  unphysical evolution is a result of the fact that we modified
  directly the transverse magnetic field on the bottom boundary in an
  unphsyical way.}, reaching $\sim 1.4$ of the potential field energy
$E_0$ (here $E_{0} = 1.2 \times 10^{33}$~erg when scaled to the
realistic value in the corona), and then keeps almost constant during
the relaxing phase with bottom boundary fixed. Very similar, the
kinetic energy first increases and later keeps on the level of
$3\times 10^{-3}$ of $E_0$. The residual of the magnetic field
increases in the first $t_s$ as we continually modified the transverse
field at the bottom boundary which drives quickly the evolution of the
field in the corona. Then it decreases to below $10^{-5}$ with a time
duration of $20~t_s$, which indicates that the magnetic field reaches
a quasi-equilibrium state.

To make the field even closer to equilibrium, we carried out a
`deeper' relaxation by running the model again, which is started with
the relaxed magnetic field obtained at $t=20~t_s$ and the initially
hydrostatic plasma. Now we reduce the kinetic viscosity to
$\nu = 0.05\Delta x^{2}/\Delta t$, i.e., an order of magnitude smaller
than the previously used one, which will let the magnetic field relax
further. Furthermore, the magnetic field at the boundary boundary is
allowed to evolve in a self-consistent way by assuming the bottom
boundary as a perfectly line-tying and fixed (i.e., $\vec v = 0$)
surface of magnetic field lines. However, such a line-tying condition
does not indicate that all magnetic field components on the
boundary are fixed, because even though the velocity $\vec v$ is given
as zero on the bottom boundary, it is not necessarily zero in the
neighboring inner points. To self-consistently update the magnetic
field, we solve the magnetic induction equation on the bottom
boundary. Slightly different from the one in the main
equations~(\ref{eq:MHD}), the induction equation at the bottom surface
is given as
\begin{equation}
\label{photo_B_equ}
  \frac{\partial \vec B}{\partial t} =  \grad \times (\vec v \times \vec B) +
  \eta_{\rm stable} \gradxy^{2} \vec B,
\end{equation}
where we added a surface diffusion term defined by using a surface
Laplace operator as
$\gradxy^{2} =\frac{\partial^{2}}{\partial
  x^{2}}+\frac{\partial^{2}}{\partial y^{2}}$
with a small resistivity for numerical stability near the PIL
$\eta_{\rm stable} = 1\times 10^{-3} e^{-B_z^2}$, since the magnetic
field often has the strongest gradient on the photosphere {around} the
main PIL. The surface induction \Eq~(\ref{photo_B_equ}) in the code is
realized by second-order difference in space and first-order forward
difference in time. Specifically, on the bottom boundary (we do not
use ghost cell), we first compute $\vec v\times \vec B$, and then use
central difference in horizontal direction and one-sided difference
(also 2nd order) in the vertical direction to compute the convection
term $\grad \times (\vec v \times \vec B)$. The surface Laplace
operator is also realized by central difference.

The curves colored in blue in \Fig~\ref{2} show the evolution of
different parameters during this relaxation phase~\footnote{{Note that the initial values of the blue curves (the deeper relaxation phase) do not equal to the values of the black curves at $t = 20 t_s$ although the deeper relaxation phase starts from that time point. This is because in the deeper relaxation process, there are three ways different from the initial relaxation one: (1) the velocity is reset to zero and the plasma is reset to hydrostatic state, thus the kinetic energy is reset to zero; (2) the viscosity is abruptly reduced by an order of magnitude; (3) the boundary condition is changed. Furthermore, not data of every time step in the run is recorded, thus small difference is shown in the magnetic energy of the initial value of the blue line (which is not exactly for the time of $t=20 t_s$) from the $t = 20 t_s$ of the black line.}}. Initially one can
see a fast decrease of the magnetic energy because the magnetic field
becomes more relaxed. As the viscosity is reduced significantly, the
kinetic energy first increases, as driven by the residual Lorentz
force of the magnetic field, to almost $5\times 10^{-3}~E_0$. Then as
the magnetic field relaxed, the kinetic energy decreases fast to
eventually less than $10^{-3}~E_0$, which is a very low level. The
residual of magnetic field of two consecutive time steps also
decreases to $10^{-5}$. These values show that the magnetic field
reaches an excellent equilibrium.

\Fig~\ref{3} shows the 3D magnetic field lines of the final relaxed
MHD equilibrium. Note that the field lines are false-colored by the
values of the force-free factor defined as
$\alpha = \vec J\cdot \vec B/B^2$, which indicates how much the field
lines are non-potential. For a force-free field, this parameter is
constant along any given field line. As can be seen, the magnetic
field is close to a perfect force-free one since the color is nearly
the same along any single field line. In the core of the
configuration, the field lines are sheared significantly along the
PIL, and thus have large values of $\alpha$ and current density. On
the other hand, the overlying field is almost current-free or
quasi-potential field with $\alpha \sim 0$, which plays the role of
strapping field that confines the inner sheared core. Such a
configuration is typical for eruption-productive
ARs~\citep[e.g.,][]{Schrijver2008a, Sun2012, Jiang2016ApJ,
  Toriumi2019, DuanA2019}.

\begin{figure}[htbp]
  \centering
  \includegraphics[width=0.48\textwidth]{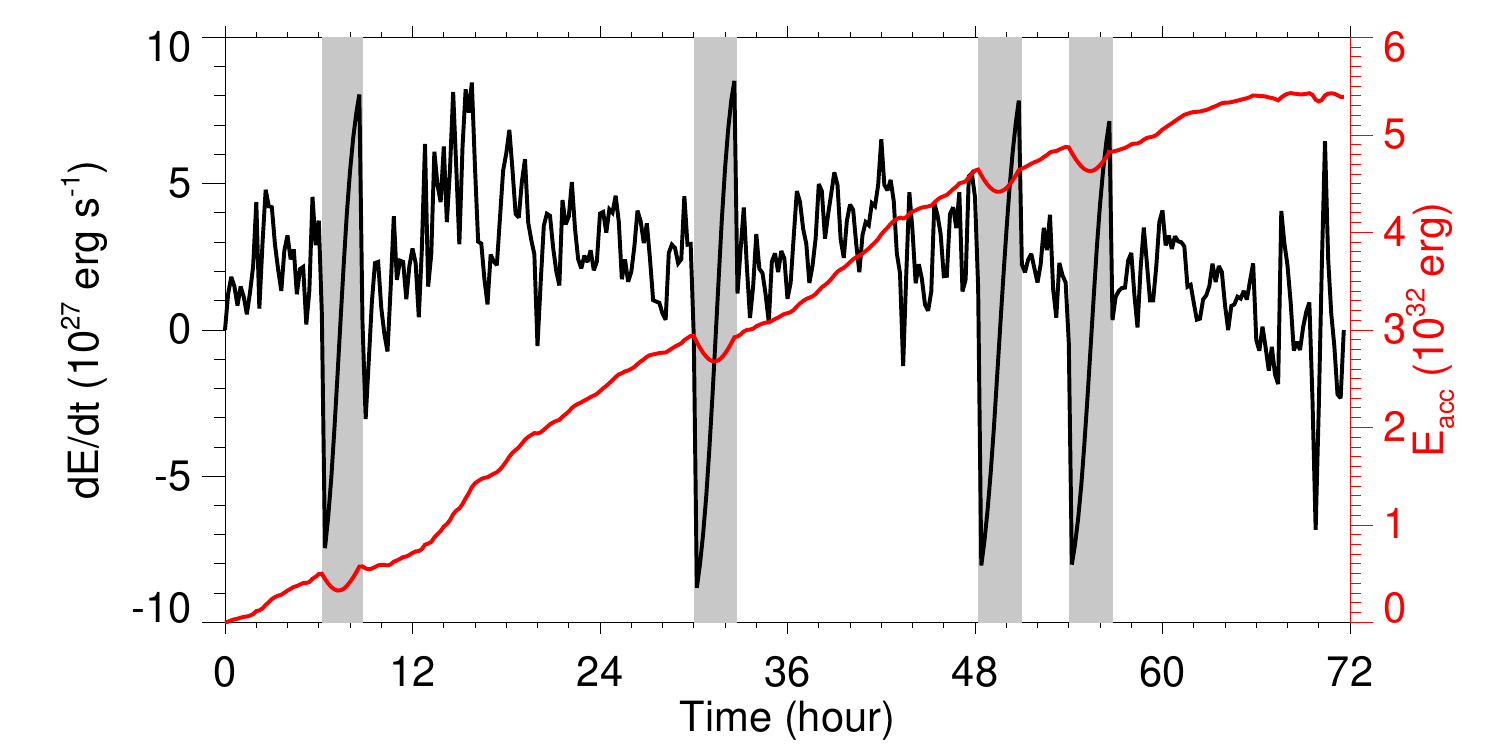}
  \caption{Evolution of Poynting flux (black line) and the accumulated
    energy (red line) which is time integration of the Poynting
    flux. The gray bars denote the data gaps of the vector
    magnetograms for which a simple linear interpolation in time is
    used to fill.}
  \label{4}
\end{figure}

\begin{figure*}[htbp]
  \centering
  \includegraphics[width=0.8\textwidth]{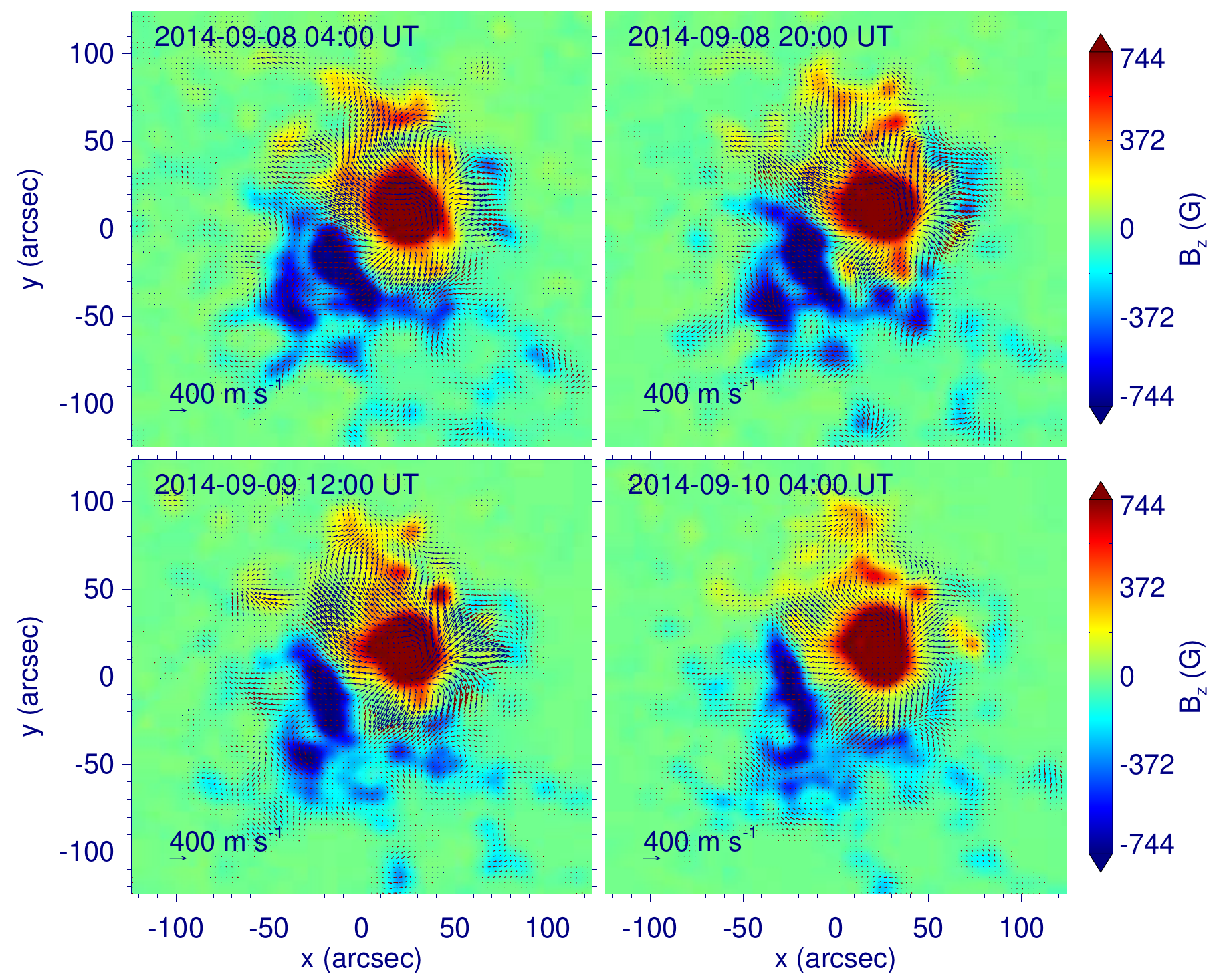}
  \caption{Four snapshots of the surface flows (the final smoothed
    version) as derived using the DAVE4VM code.}
  \label{5}
\end{figure*}

\section{Derive the surface flow field}
\label{flow}

Based on the time sequence of vector magnetograms, it is
straightforward to derive the surface velocity by employing the
DARE4VM code developed by~\citet{Schuck2008}. The differential affine
velocity estimator (DAVE) was first developed for estimating
velocities from line-of-sight magnetograms, and was then modified to
directly incorporate horizontal magnetic fields to produce a
differential affine velocity estimator for vector magnetograms
(DAVE4VM). We use the SHARP data with cadence of 12 minutes and pixel
size of 1~arcsec (by rebinning the original data with pixel size of
0.5~arcsec), and first of all, we fill the data gap using a linear
interpolation in the time domain to generate a complete time series of
three days from 00:00~UT on 2014 September 8 to 24:00~UT on September
10 with cadence of 12 minutes. Then we input the time series of vector
magnetogram (rebinned as 1~arcsec per pixel) in the DAVE4VM code. A
key parameter needed by the DAVE4VM code is the window size, and here
we use 19~pixels, following~\citet{Liu2012} and
\citet{LiuC2014}. After obtaining the surface velocity, we first reset
those in the weak-field region (with total magnetic field strength
below 100~G) as zero, because there are large errors and unresolved
scales in these regions. \Fig~\ref{4} shows evolution of the Poynting
flux ${\rm d}E/{\rm d}t$, which is defined by
\begin{equation}\label{poy}
  \frac{{\rm d} E}{{\rm d} t} = \frac{1}{\mu_0}\int_S \left[(B_{x}^{2}+ B_{y}^{2})v_{z} - ( B_{x}v_{x}+B_{y}v_{y} )B_{z}\right] {\rm d}x{\rm d}y,
\end{equation}
where $S$ is the photospheric surface, and its time accumulation
$E_{\rm acc}$, as computed by using the surface flow $(v_x, v_y, v_z)$
and the magnetic field. It can be seen that, except the data gap
intervals\footnote{During the data gaps the Poynting flux is found to
  become negative adruptly, which might be a result of our simple
  linear interpolation in filling the gaps of the magnetograms. More
  optimized method for filling the data gaps will be considered to
  recover a more consistent evolution of Poynting flux.}, the magnetic
energy is continually injected in the corona through the photosphere,
and in the three days, it gains $\sim 5\times 10^{32}$~erg.

Before being input into the model, the flow data are also needed to be
smoothed. We smoothed the time series of flow maps in both the time
and space domains, with a Gaussian FWHM of 6 hours (i.e., 30 time
snapshots) and 8~arcsec, respectively, which is finally input to the
data-driven model. \Fig~\ref{5} shows 4 snapshots of the surface
velocity after smoothing. The speed of the flow is generally a few
hundreds of meters per second and the main feature is a clear and
persistent rotation of the main sunspot. Note that during the three
days the basic configuration of the photospheric magnetic flux
distribution is rather similar with only somewhat dispersion. So the
magnetic energy injection should come mainly from the transverse
rotational flows. In addition to the rotational flow, we can see very
evident diverging flow existing persistently near the boundary of the
sunspot.

\begin{figure}[htbp]
  \centering
  \includegraphics[width=0.45\textwidth]{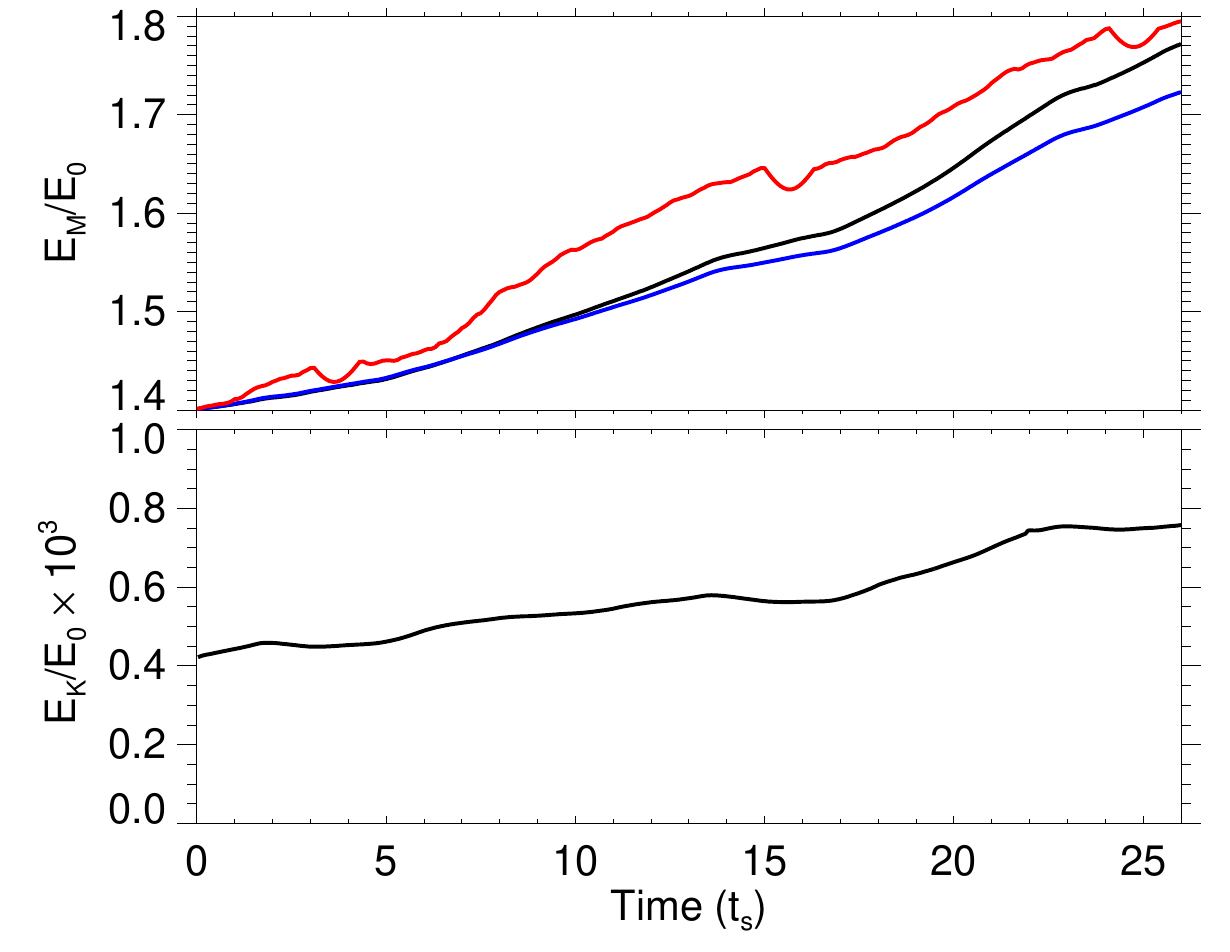}
  \caption{Magnetic energy (top) and kinetic energy (bottom) evolution
    of the MHD system as driven by the photospheric flows. The black lines represent the values of the volume-integrated energies. The blue
    line in the top panel shows the time integration of the total
    Poynting flux on the bottom surface from the simulation. The red
    line shows the accumulated magnetic energy from directly the
    observation data, i.e., the same one shown in \Fig~\ref{4}.}
  \label{6}
\end{figure}

\begin{figure*}[htbp]
  \centering
  \includegraphics[width=0.9\textwidth]{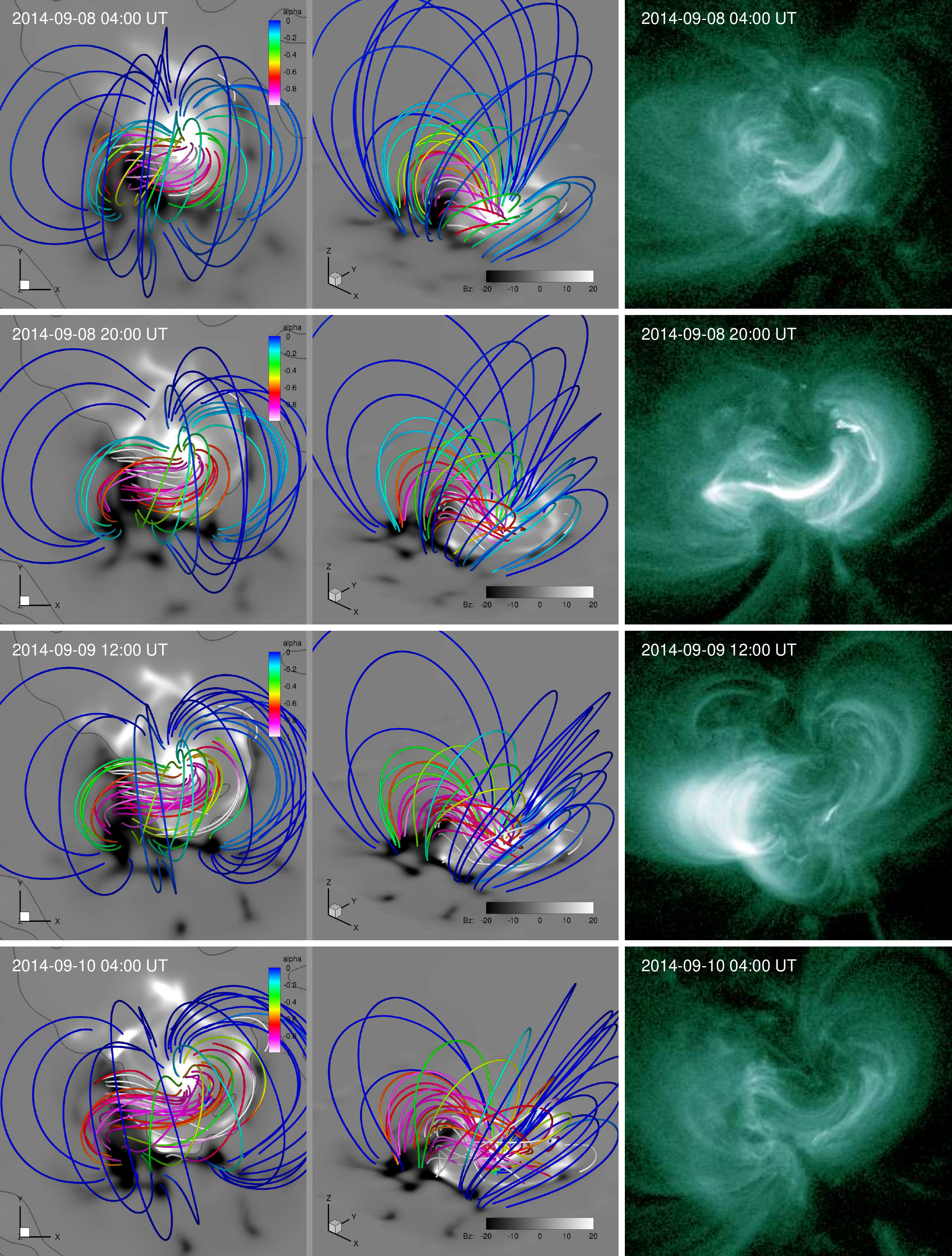}
  \caption{Evolution of the magnetic field lines (shown in different
    view angles in the left and middle columns) and their comparison
    with observed coronal loops observed in EUV wavelength of 94~{\AA}
    by SDO/AIA (right column). The field lines are shown in the same
    format in \Fig~\ref{3}. An animation of the magnetic field
    evolution is attached for this figure.}
  \label{7}
\end{figure*}

\begin{figure}[htbp]
  \centering
  \includegraphics[width=0.49\textwidth]{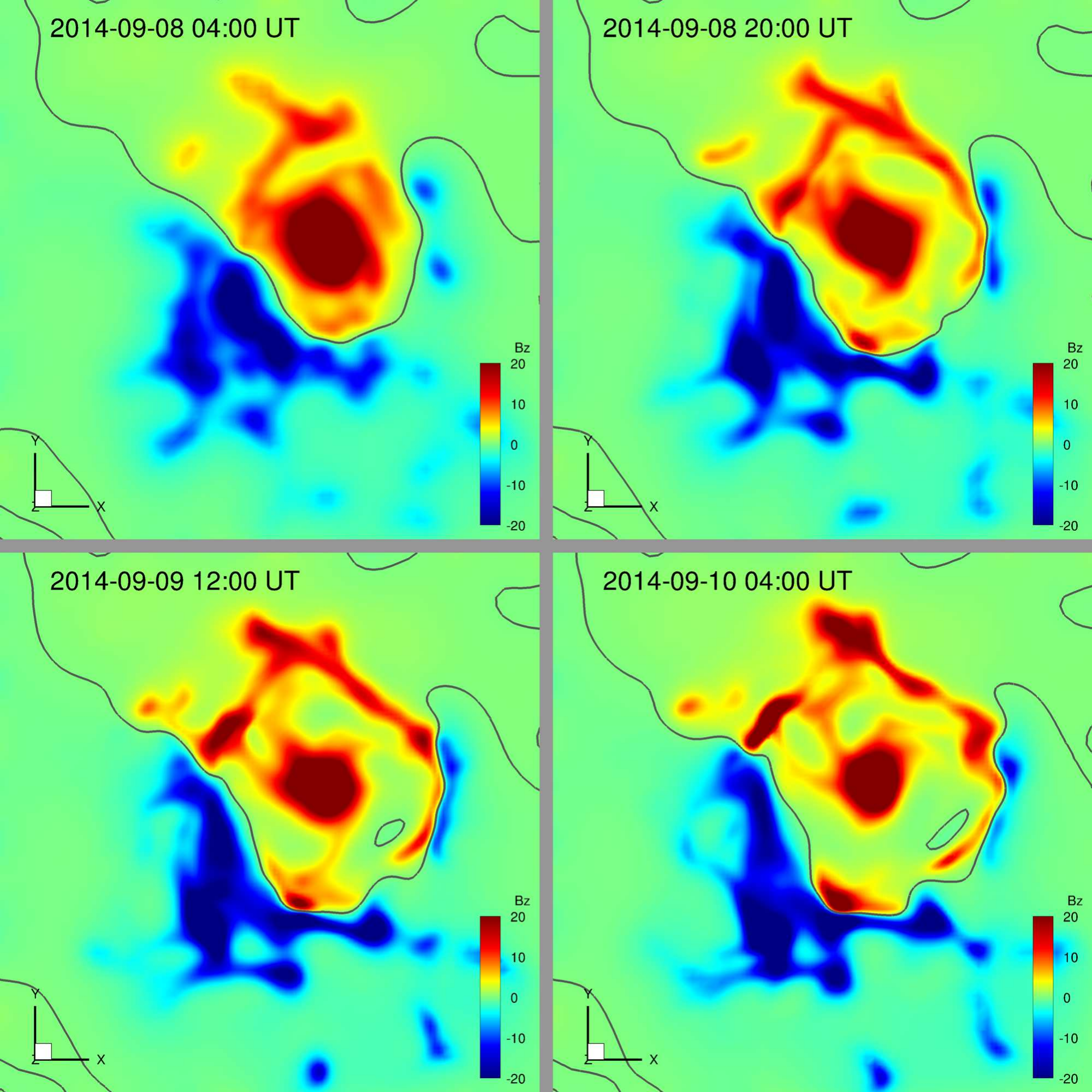}
  \caption{Evolution of magnetic flux distribution on the bottom
    surface in the simulation, shown for 4 snapshots {of the same times shown
    in \Fig~\ref{5}}.}
  \label{8}
\end{figure}

\section{Data-driven simulation}
\label{evol}

We input the surface velocity at the bottom boundary to drive the
evolution of the model, by starting the simulation from the solution
obtained from the time point of $t=58$ in the relaxation phase (see \Fig~\ref{2}) as
described in Section~\ref{initial}. Here to save computing time, the
cadence of the input flow maps, which is originally 12 min, was
increased by $68.6$ times when inputting into the MHD model. As a
result, an unit of time in the simulation, $t_s$, corresponds to
actually $t_s \times 68.6=7200$~s, i.e., 2 hours, in the HMI
data. Compressing of the time in HMI data is justified by the fact
that the speed of photospheric flows is often a few
$100$~m~s$^{-1}$. So in our model settings, the evolution speed of the
boundary field, even enhanced by a factor of $68.6$, is still smaller
by two orders of magnitude than the coronal Alfv{\'e}n speed (on the
order of $10^3$~km~s$^{-1}$), and the quick reaction of the coronal
field to the slow bottom changes should not be affected. The
implementation of the bottom boundary conditions is the same as that
for the deeper relaxation phase described in
Section~\ref{initial}. That is, on the bottom surface, we solved the
equation~(\ref{photo_B_equ}) to update all the three components of
magnetic field with the flow field prescribed by those derived in
Section~\ref{flow}, while the plasma density and temperature are
simply fixed. In the driven-evolution phase, the kinetic viscosity is
also used as the smaller one $\nu = 0.05\Delta x^{2}/\Delta t$, which
corresponds to a Reynolds number of $10$ for the length of a grid cell
$\Delta x$.

We show an approximately two-day (a duration of $26~t_s$ or $52$ hours
in reality) evolution of the MHD system as an example, while the
further evolution associated with an eruption and the physical
mechanisms will be left for future study. \Fig~\ref{6} presents the
global energy evolutions. It can be seen the magnetic energy (the
black line in the top panel) increases monotonously as driven by the
surface flows. By the end of our simulation, it reaches approximately
$1.8$ times of the initial potential energy, and thus the total
magnetic energy obtained in this driving phase is
$(1.8-1.4) \times E_0 \approx 4.8 \times 10^{32}$~erg. On the other
hand, the kinetic energy (bottom panel of \Fig~\ref{6}) keeps below
the level of $1\times 10^{-3} E_0$ with a mild increase, which
indicates that the system remains a stable, quasi-static evolution. In
the top panel of \Fig~\ref{6} we also plots the time integration of
total Poynting flux (the blue line), using the magnetic field and the
flow field on the bottom boundary of the simulation, which is the
energy injected into the volume from the bottom boundary through the
surface flow. If our boundary condition is accurately implemented, the
energy injected from the bottom surface should match the magnetic
energy obtained in the computational volume, since other energies are
negligible if compared with the magnetic energy. As can be seen, the
trend of magnetic energy evolution (the black line) matches rather
well that of the energy input by the surface flow (the blue line),
{albeit a small numerical error that increases slowly the total
magnetic energy, which is also seen in the relaxation phase as shown
in \Fig~\ref{2}}. It is worthy noting that the magnetic energy
evolution is also in good agreement with the accumulated magnetic
energy as derived from directly the observation data (the red line, {which is the same value shown in \Fig~\ref{4} but multiplied by a factor of $0.05^2$ because the field input to the model is multiplied by $0.05$}) as
calculated in Section~\ref{flow}, which suggests that our data-driven
model can reliably reproduce the magnetic energy injection into the
corona through the photospheric motions. {The small difference between the observation-derived and simulated energies might be due to the smoothing of the velocity since it filters out the small-scale flows that also contribute to the total Poynting flux.}

\Fig~\ref{7} (and its attached animation) shows the 3D magnetic field
lines and their evolution in comparison with SDO/AIA image of coronal
loops in the 94~{\AA} wavelength which highlights the hot, core
coronal loops in the AR. Overall, we can see a slow stressing of the
field lines mainly through the sunspot rotational flow. It renders the
core field lines to form a more and more coherent S shape, which
resembles the observed sigmoid structure in the core of the AR. The
increase of non-potentiality can also be seen from the increase of the
force-free factor $\alpha$ in the core region. \Fig~\ref{8} shows
evolution of the surface magnetic flux distribution (see also the
animation attached to \Fig~\ref{7}). {The four snapshots are taken
for the same times given in \Fig~\ref{5}, and thus comparing \Fig~\ref{8} with \Fig~\ref{5}
shows the difference between the simulated magnetograms and the observed ones.}
In addition to the rotation of
the main sunspot, an evident feature is the enhancement of the field
strength along the PIL. This is owing to the divergence flow from the
edge of the sunspot, which continually convect the magnetic flux to
the PIL. Such pileup of magnetic flux near the PIL, however, is not
seen in the observed magnetograms (\Fig~\ref{5}), which rather show
field decaying. Such decaying of magnetic flux is likely due to the
global turbulent diffusion of photospheric magnetic field by granular
and supergranular convection~\citep{Wang1989} and other small-scale
turbulence and flux cancellations in the photosphere, which is not
being recovered by the DAVE4VM code and thus not reproduced in our
simulation.

\section{Conclusions}
\label{concl}

This paper is devoted to the description of a new approach of
data-driven modelling of solar AR magnetic field evolution,
{in which, we have for the first time utilized directly the photospheric velocity field from
 DAVE4VM to drive a full-MHD model.}
To setup
the initial conditions, we used a special MHD relaxation approach with
sufficiently small kinetic viscosity to construct a true MHD
equilibrium based on a single vector magnetogram. Then we derived the
photospheric surface flows from a time series of observed magentograms
based on the DAVE4VM method. The surface flows were finally inputted,
again in time sequence, to the bottom boundary of the MHD model to
self-consistently update the magnetic field at every time step, which
is implemented by solving directly the magnetic induction equation at the
bottom boundary  using finite difference method.

We applied this data-driven model to study the magnetic field
evolution of AR~12158 with SDO/HMI vector magnetograms. The initial
MHD equilibrium is calculated using magnetogram observed for 00:00~UT
on 2014 September 8, and a two-day duration of the AR evolution is
then simulated using the data-driven MHD model. Overall, the evolution
is characterized by a slow stress of the field lines mainly through
the rotational flow of the AR's leading sunspot, which makes the core
field lines to forms an coherent S shape consistent with the sigmoid
structure as seen in the AIA images. Such evolution proceeds in a
quasi-static way since the kinetic energy of the system remains less
than its magnetic energy by three orders of magnitude, while the
magnetic energy increases monotonously as driven by the surface flow,
and reaches approximately double of the initial potential energy by
the end of the simulation. The magnetic energy obtained in the
simulation during the surface driving phase matches closely the
accumulated magnetic energy as calculated directly from the original
vector magnetogram with the DAVE4VM derived flow field.

With the surface flow specified at the bottom boundary, the magnetic
field can be updated self-consistently by solving the induction
equation at the surface boundary. However, our simulation shows that
discrepancy arises between the simulated magnetic field at the bottom
surface with the original magnetograms. That is, in the simulation,
magnetic flux are significantly enhanced along the PIL, owing to the
divergence flow from the edge of the sunspot, whereas in the observed
magnetograms such pileup of magnetic flux near the PIL is not seen,
which rather shows field decaying. {In the next step, we will
  try to use some ad-hoc flux cancellation near the PIL, for which a
  straightforward way is to increase the value of $\eta_{\rm stable}$
  in equation~(\ref{photo_B_equ}), such that the flux diffusion speed
  is comparable to that of the flux pileup.}  In the future, we will
consider to include the global diffusion of photospheric magnetic
field by granular and supergranular convection to simulate more
realistically the magnetic field evolution. {On the other hand,
  the discrepancy might result from errors in the vector magnetograms,
  since these errors can introduce spurious flows with the DAVE4VM
  code and thus influences our simulation. To elucidate this, we will
  test our model with error-free magnetograms from recent convective
  flux-emergence
  simulations~\citep[e.g.,][]{ChenF2017,Cheung2019,Toriumi2019S} as
  the ground-truth data.}

{Our ultimate purpose is to use the model} to
study how the coronal field, as driven by the slow photospheric motions,
reaches a unstable state and runs into eruption. From this, we will be able to
investigate in details the topology of the evolving magnetic field leading to the eruption,
to see whether it forms a magnetic flux rope and becomes ideally unstable~\citep[e.g.,][]{Kliem2006, Aulanier2010}, or, a simply sheared
 arcade with a interally-formed current sheet to trigger flare reconnection~\citep{Moore2001}, which
would be helpful for resolving the long-standing debates on the triggering mechanism of
solar eruptions~\citep{ChenP2011}.

\acknowledgments This work is jointly supported by National Natural
Science Foundation of China (NSFC 41822404, 41731067, 41574170,
41531073), the Fundamental Research Funds for the Central Universities
(Grant No.HIT.BRETIV.201901) and Shenzhen Technology Project
JCYJ20190806142609035. Data from observations are courtesy of NASA
SDO. The computational work was carried out on TianHe-1(A), National
Supercomputer Center in Tianjin, China. {We are grateful to the reviewers for
helping to improve the paper.}


\begin{thebibliography}{44}
\expandafter\ifx\csname natexlab\endcsname\relax\def\natexlab#1{#1}\fi

\bibitem[{{Aulanier} {et~al.}(2010){Aulanier}, {T{\"o}r{\"o}k}, {D{\'e}moulin},
  \& {DeLuca}}]{Aulanier2010}
{Aulanier}, G., {T{\"o}r{\"o}k}, T., {D{\'e}moulin}, P., \& {DeLuca}, E.~E.
  2010, \apj, 708, 314

\bibitem[{{Chen} {et~al.}(2017){Chen}, {Rempel}, \& {Fan}}]{ChenF2017}
{Chen}, F., {Rempel}, M., \& {Fan}, Y. 2017, \apj, 846, 149

\bibitem[{{Chen}(2011)}]{ChenP2011}
{Chen}, P.~F. 2011, Living Reviews in Solar Physics, 8, 1

\bibitem[{Cheung \& DeRosa(2012)}]{Cheung2012}
Cheung, M. C.~M. \& DeRosa, M.~L. 2012, The Astrophysical Journal, 757, 147

\bibitem[{Cheung {et~al.}(2019)Cheung, Rempel, Chintzoglou, Chen, Testa,
  Mart{\'i}nez-Sykora, Sainz~Dalda, DeRosa, Malanushenko, Hansteen, De~Pontieu,
  Carlsson, Gudiksen, \& McIntosh}]{Cheung2019}
Cheung, M. C.~M., Rempel, M., Chintzoglou, G., Chen, F., Testa, P.,
  Mart{\'i}nez-Sykora, J., Sainz~Dalda, A., DeRosa, M.~L., Malanushenko, A.,
  Hansteen, V., De~Pontieu, B., Carlsson, M., Gudiksen, B., \& McIntosh, S.~W.
  2019, Nature Astronomy, 3, 160

\bibitem[{Duan {et~al.}(2019)Duan, Jiang, He, Feng, Zou, \& Cui}]{DuanA2019}
Duan, A., Jiang, C., He, W., Feng, X., Zou, P., \& Cui, J. 2019, The
  Astrophysical Journal, 884, 73

\bibitem[{Duan {et~al.}(2017)Duan, Jiang, Hu, Zhang, Gary, Wu, \&
  Cao}]{DuanA2017}
Duan, A., Jiang, C., Hu, Q., Zhang, H., Gary, G.~A., Wu, S.~T., \& Cao, J.
  2017, The Astrophysical Journal, 842, 119

\bibitem[{Feng {et~al.}(2012)Feng, Jiang, Xiang, Zhao, \& Wu}]{FengX2012}
Feng, X., Jiang, C., Xiang, C., Zhao, X., \& Wu, S.~T. 2012, The Astrophysical
  Journal, 758, 62

\bibitem[{Guo {et~al.}(2019)Guo, Xia, Keppens, Ding, \& Chen}]{GuoY2019}
Guo, Y., Xia, C., Keppens, R., Ding, M.~D., \& Chen, P.~F. 2019, The
  Astrophysical Journal, 870, L21

\bibitem[{{Hayashi}(2005)}]{Hayashi2005}
{Hayashi}, K. 2005, \apjs, 161, 480

\bibitem[{Hayashi {et~al.}(2019)Hayashi, Feng, Xiong, \& Jiang}]{Hayashi2019}
Hayashi, K., Feng, X., Xiong, M., \& Jiang, C. 2019, The Astrophysical Journal,
  871, L28

\bibitem[{{Inoue}(2016)}]{Inoue2016Review}
{Inoue}, S. 2016, Progress in Earth and Planetary Science, 3, 19

\bibitem[{Inoue {et~al.}(2018)Inoue, Kusano, B{\"u}chner, \&
  Sk{\'a}la}]{Inoue2018}
Inoue, S., Kusano, K., B{\"u}chner, J., \& Sk{\'a}la, J. 2018, Nature
  communications, 9, 174

\bibitem[{{Jiang} \& {Feng}(2013)}]{Jiang2013NLFFF}
{Jiang}, C. \& {Feng}, X. 2013, \apj, 769, 144

\bibitem[{{Jiang} \& {Feng}(2014)}]{Jiang2014Prep}
---. 2014, \solphys, 289, 63

\bibitem[{Jiang \& Toriumi(2020)}]{JiangC2020}
Jiang, C. \& Toriumi, S. 2020, The Astrophysical Journal, 903, 11

\bibitem[{Jiang {et~al.}(2016)Jiang, Wu, Yurchyshyn, Wang, Feng, \&
  Hu}]{Jiang2016ApJ}
Jiang, C., Wu, S.~T., Yurchyshyn, V.~B., Wang, H., Feng, X., \& Hu, Q. 2016,
  \apj, 828, 62

\bibitem[{Jiang {et~al.}(2010)Jiang, Feng, Zhang, \& Zhong}]{Jiang2010}
Jiang, C.~W., Feng, X.~S., Zhang, J., \& Zhong, D.~K. 2010, \solphys, 267, 463

\bibitem[{{Jiang} {et~al.}(2016){Jiang}, {Wu}, {Feng}, \& {Hu}}]{Jiang2016NC}
{Jiang}, C.~W., {Wu}, S.~T., {Feng}, X.~S., \& {Hu}, Q. 2016, Nature Comm., 7,
  11522

\bibitem[{{Kliem} \& {T{\"o}r{\"o}k}(2006)}]{Kliem2006}
{Kliem}, B. \& {T{\"o}r{\"o}k}, T. 2006, Physical Review Letters, 96, 255002

\bibitem[{Leake {et~al.}(2017)Leake, Linton, \& Schuck}]{Leake2017}
Leake, J.~E., Linton, M.~G., \& Schuck, P.~W. 2017, The Astrophysical Journal,
  838, 113, arXiv: 1702.06808

\bibitem[{Lee \& Magara(2018)}]{Lee2018}
Lee, H. \& Magara, T. 2018, The Astrophysical Journal, 859, 132

\bibitem[{Liu {et~al.}(2014)Liu, Deng, Lee, Wiegelmann, Jiang, Dennis, Su,
  Donea, \& Wang}]{LiuC2014}
Liu, C., Deng, N., Lee, J., Wiegelmann, T., Jiang, C., Dennis, B., Su, Y.,
  Donea, A., \& Wang, H. 2014, \apj, 795, 128

\bibitem[{{Liu} {et~al.}(2012){Liu}, {Zhao}, \& {Schuck}}]{Liu2012}
{Liu}, Y., {Zhao}, J., \& {Schuck}, P.~W. 2012, \solphys, 195

\bibitem[{{Moore} {et~al.}(2001){Moore}, {Sterling}, {Hudson}, \&
  {Lemen}}]{Moore2001}
{Moore}, R.~L., {Sterling}, A.~C., {Hudson}, H.~S., \& {Lemen}, J.~R. 2001,
  \apj, 552, 833

\bibitem[{{Nakagawa}(1980)}]{Nakagawa1980}
{Nakagawa}, Y. 1980, \apj, 240, 275

\bibitem[{{Nakagawa} {et~al.}(1987){Nakagawa}, {Hu}, \& {Wu}}]{Nakagawa1987}
{Nakagawa}, Y., {Hu}, Y.~Q., \& {Wu}, S.~T. 1987, \aap, 179, 354

\bibitem[{Pomoell {et~al.}(2019)Pomoell, Lumme, \& Kilpua}]{Pomoell2019}
Pomoell, J., Lumme, E., \& Kilpua, E. 2019, Solar Physics, 294, 41

\bibitem[{{Priest} \& {Forbes}(2002)}]{Priest2002}
{Priest}, E.~R. \& {Forbes}, T.~G. 2002, \aapr, 10, 313

\bibitem[{{Schrijver} {et~al.}(2008){Schrijver}, {DeRosa}, {Metcalf}, {Barnes},
  {Lites}, {Tarbell}, {McTiernan}, {Valori}, {Wiegelmann}, {Wheatland},
  {Amari}, {Aulanier}, {D{\'e}moulin}, {Fuhrmann}, {Kusano}, {R{\'e}gnier}, \&
  {Thalmann}}]{Schrijver2008a}
{Schrijver}, C.~J., {DeRosa}, M.~L., {Metcalf}, T., {Barnes}, G., {Lites}, B.,
  {Tarbell}, T., {McTiernan}, J., {Valori}, G., {Wiegelmann}, T., {Wheatland},
  M.~S., {Amari}, T., {Aulanier}, G., {D{\'e}moulin}, P., {Fuhrmann}, M.,
  {Kusano}, K., {R{\'e}gnier}, S., \& {Thalmann}, J.~K. 2008, \apj, 675, 1637

\bibitem[{Schuck(2008)}]{Schuck2008}
Schuck, P.~W. 2008, The Astrophysical Journal, 683, 1134

\bibitem[{{Sun} {et~al.}(2012){Sun}, {Hoeksema}, {Liu}, {Wiegelmann},
  {Hayashi}, {Chen}, \& {Thalmann}}]{Sun2012}
{Sun}, X., {Hoeksema}, J.~T., {Liu}, Y., {Wiegelmann}, T., {Hayashi}, K.,
  {Chen}, Q., \& {Thalmann}, J. 2012, \apj, 748, 77

\bibitem[{Toriumi \& Hotta(2019)}]{Toriumi2019S}
Toriumi, S. \& Hotta, H. 2019, The Astrophysical Journal, 886, L21

\bibitem[{Toriumi {et~al.}(2020)Toriumi, Takasao, Cheung, Jiang, Guo, Hayashi,
  \& Inoue}]{Toriumi2020}
Toriumi, S., Takasao, S., Cheung, M. C.~M., Jiang, C., Guo, Y., Hayashi, K., \&
  Inoue, S. 2020, The Astrophysical Journal, 890, 103

\bibitem[{Toriumi \& Wang(2019)}]{Toriumi2019}
Toriumi, S. \& Wang, H. 2019, Living Reviews in Solar Physics, 16, 3

\bibitem[{Vemareddy {et~al.}(2016)Vemareddy, Cheng, \&
  Ravindra}]{Vemareddy2016}
Vemareddy, P., Cheng, X., \& Ravindra, B. 2016, The Astrophysical Journal, 829,
  24

\bibitem[{{Wang} {et~al.}(1989){Wang}, {Nash}, \& {Sheeley}}]{Wang1989}
{Wang}, Y., {Nash}, A.~G., \& {Sheeley}, Jr., N.~R. 1989, Science, 245, 712

\bibitem[{{Wiegelmann}(2004)}]{Wiegelmann2004}
{Wiegelmann}, T. 2004, \solphys, 219, 87

\bibitem[{{Wiegelmann} \& {Sakurai}(2012)}]{Wiegelmann2012solar}
{Wiegelmann}, T. \& {Sakurai}, T. 2012, Living Reviews in Solar Physics, 9, 5

\bibitem[{{Wu} {et~al.}(2006){Wu}, {Wang}, {Liu}, \& {Hoeksema}}]{Wu2006}
{Wu}, S.~T., {Wang}, A.~H., {Liu}, Y., \& {Hoeksema}, J.~T. 2006, \apj, 652,
  800

\bibitem[{Wu \& Wang(1987)}]{Wu1987}
Wu, S.~T. \& Wang, J.~F. 1987, Computer Methods in Applied Mechanics and
  Engineering, 64, 267

\bibitem[{{Yang} {et~al.}(1986){Yang}, {Sturrock}, \& {Antiochos}}]{Yang1986}
{Yang}, W.~H., {Sturrock}, P.~A., \& {Antiochos}, S.~K. 1986, \apj, 309, 383

\bibitem[{Zhao {et~al.}(2016)Zhao, Gilchrist, Aulanier, Schmieder, Pariat, \&
  Li}]{ZhaoJ2016}
Zhao, J., Gilchrist, S.~A., Aulanier, G., Schmieder, B., Pariat, E., \& Li, H.
  2016, The Astrophysical Journal, 823, 62, arXiv: 1603.07563

\bibitem[{Zhou {et~al.}(2016)Zhou, Zhang, \& Wang}]{ZhouG2016}
Zhou, G.~P., Zhang, J., \& Wang, J.~X. 2016, The Astrophysical Journal, 823,
  L19

\end{thebibliography}

\end{document}